\documentclass[showpacs,aps,prl,twocolumn,longcitation]{revtex4-1}
\usepackage{graphicx}
\usepackage{bm}
\usepackage{color}

\usepackage[utf8]{inputenc}
\usepackage{amsmath}
\usepackage{amssymb}
\usepackage{enumerate}
\usepackage{xspace}
\usepackage{mathrsfs}
\usepackage{mathptmx}

\newcommand{\kc}{\ensuremath{k_{\mathrm{c}}}\xspace}

\newcommand{\smJij}{\ensuremath{J_{ij}}\xspace}
\newcommand{\smmu}{\ensuremath{\mu_{i}}\xspace}
\newcommand{\smH}{\ensuremath{\mathbf{B}}\xspace}

\newcommand{\sms}{\ensuremath{\mathbf{S}}\xspace}
\newcommand{\vampire}{\textsc{vampire} }
\newcommand{\Tc}{\ensuremath{T_{\mathrm{c}}}\xspace}

\newcommand{\muB}{\ensuremath{\mu_{\mathrm{B}}}\xspace}

\newcommand{\Te}{\ensuremath{T_{\mathrm{e}}'}\xspace}

\begin{document}

\title{Massively parallel atomistic simulation of ultrafast thermal spin dynamics of a permalloy vortex}
\author{Daniel~Meilak}
\affiliation{Department of Physics, The University of York, York, YO10 5DD, UK}
\author{Sarah~Jenkins}
\affiliation{Department of Physics, The University of York, York, YO10 5DD, UK}
\author{Rory~Pond}
\affiliation{Department of Physics, The University of York, York, YO10 5DD, UK}
\author{Richard~F.~L.~Evans}
\email{richard.evans@york.ac.uk}
\affiliation{Department of Physics, The University of York, York, YO10 5DD, UK}
\begin{abstract}
Ultrafast magnetization dynamics probes the most fundamental properties of magnetic materials, exploring questions about the fundamental interactions responsible for magnetic phenomena. Thermal effects are known to be extremely important for laser-induced dynamics in metallic systems, but the dynamics of topological magnetic structures are little understood. Here we apply a massively parallel atomistic spin dynamics simulation to study the response of a permalloy vortex to a 50 fs laser pulse. We find that macroscopically the short timescale dynamics are indistinguishable from the bulk, but that strong edge spin waves lead to a complex time evolution of the magnetic structure and long-lived oscillations on the nanosecond timescale. In the near future such simulations will provide unprecedented insight into the dynamics of magnetic materials and devices beyond the approximations of continuum micromagnetics.
\end{abstract}

\date{\today}
\maketitle

Since the pioneering experiment of Beaurepaire \textit{et al}~\cite{BeaurepairePRL1996} demonstrating sub-picosecond demagnetization of Ni in response to laser excitation, ultrafast magnetism has become one of the most active areas of research in the field. This has lead to a wide range of phenomena including inertia driven spin excitations\cite{KimelAOS2005}, helicity-dependent all-optical switching~\cite{StanciuAOS2007,Mangin2014}, and thermally induced switching~\cite{RaduNature2011,OstlerNatCom2012}. A theoretical understanding of the physical origins of the diversity of phenomena is at an early stage, but it is clear that elevated temperatures and transverse spin fluctuations play an important if not dominating role in laser induced magnetisation dynamics. Most models and interpretations of ultrafast magnetisation dynamics have relied on a single domain approximation, where the magnetisation is assumed to be approximately uniform in the sample. However, this approximation breaks down in the case of materials with topological magnetic structures such as magnetic domains, vortices, Bloch points or Skyrmions. It is an open question how these inhomogeneous structures may affect the response of a material to ultrafast laser excitation, though initial experimental studies have suggested a range of complex and statistical dynamic effects \cite{LeGuyaderAPL2012,FuSciAdv2018}.

Permalloy (Ni$_{80}$Fe$_{20}$) is an ideal material for studying magnetisation dynamics experimentally due to its intrinsically low magnetocrystalline anisotropy and compatibility with nanoscale patterning. This combination of relatively high magnetic moment and low anisotropy also led to theoretical understanding of topological magnetic vortex structures in nanodots \cite{ShinjoScience2000,ScholzMAGPAR2003}. Despite its relative simplicity, there are open questions regarding the dynamics and temperature dependent properties of magnetic vortices which are not properly addressed with continuum micromagnetic models. In particular, high temperatures and rapid spatial variations of the magnetization and ultrafast dynamics are not accessible to conventional micromagnetic models \cite{ScholzMAGPAR2003,WysinPRB2012,FuSciAdv2018,IacoccaNatComm2019}, though Landau-Lifshitz-Bloch models may describe the essential thermodynamic behaviour\cite{AtxitiaPRB2010, HinzkePRB2015, LebeckiPRB2014}.

With the recent development of massively parallel atomistic spin dynamics codes~\cite{EvansVMPR2013,UPPASD} and wider availability of High Performance Computing resources, it is now possible to simulate a sufficiently large nanodot with atomistic resolution such that a magnetic vortex is the natural ground state. Unlike previous micromagnetic simulations\cite{ScholzMAGPAR2003,WysinPRB2012,FuSciAdv2018}, thermal effects are naturally included and high energy spin waves are simulated directly. To simulate the thermal properties of Permalloy under the influence of temperature and laser pulses we construct an atomistic spin model\cite{EvansVMPR2013} writing a spin Hamiltonian with Heisenberg exchange of the form 
\begin{equation}
    \mathscr{H} = -\sum_{i < j} \smJij \sms_i \cdot \sms_j +\frac{\kc}{2} \left(S_x^4 + S_y^4 + S_z^4 \right) - \sum_i \smmu \sms_i \cdot \smH_{\mathrm{dip}}
    \label{eq:ham}
\end{equation}
where $\sms_i$ and $\sms_j$ are unit spin directions at local sites $i$ and neighbouring sites $j$ respectively, $\smJij = 3.78 \times 10^{-21}$ J/link is the exchange interaction between nearest neighbouring spins, $\kc = 3.355 \times 10^{-26}$ J/atom is the cubic anisotropy constant, $\smmu$ is the atomic spin moment at lattice site $i$ and $\smH_{\mathrm{dip}}$ is the dipole field originating from the sample. The spins are constructed on a face-centred cubic lattice with local moments of $\mu_{\mathrm{Fe}} = 2.9 \muB$ and $\mu_{\mathrm{Ni}} = 0.62 \muB$ for Fe and Ni sites respectively\cite{RaduSPIN2015}. 

\begin{figure}[tb]
\center
\includegraphics[width=8.3cm, trim=0 0 0 0]{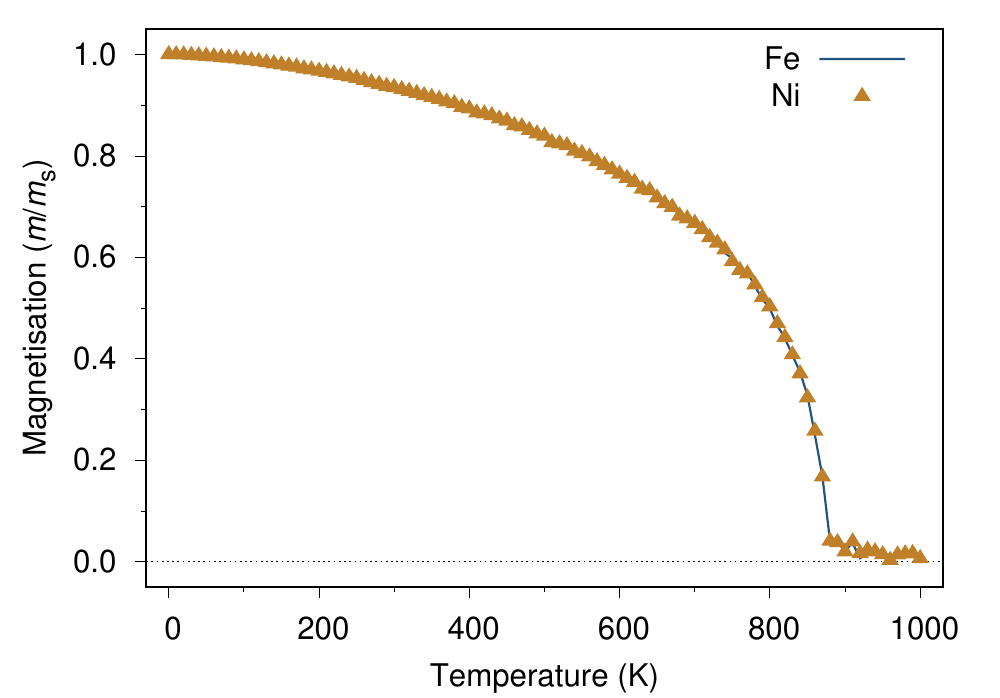}
\caption{(Color online) Simulated temperature dependent magnetization for Ni$_{80}$Fe$_{20}$ Permalloy using Monte Carlo simulations. The Fe and Ni sublatttices show full ferromagnetic alignment at equilibrium due to the strong ferromagnetic exchange.}
\label{fig:mvsT}
\end{figure}

To model the equilibrium temperature dependent properties of Permalloy we simulate the temperature dependent magnetization of a (10 nm)$^3$ with periodic boundary conditions shown in Fig.~\ref{fig:mvsT} using a Monte Carlo Metropolis simulation with an adaptive move\cite{AlzateCardonaJPCM2019} as implemented in the \vampire software package \cite{EvansVMPR2013,vampire-url}. To account for the quantum nature of the thermal fluctuations we apply spin temperature rescaling\cite{EvansPRB2015} with an exponent $\eta = 1.63$ fitted from experimental data for bulk Permalloy\cite{LuoJMMM2015}, giving quantitative agreement with experimental data for the macroscopic temperature dependent properties and Curie temperature. At thermal equilibrium the Fe and Ni sublattices in the Permalloy are strongly exchange coupled and therefore have the same temperature dependence for their intrinsic magnetic properties.
While the equilibrium properties can be effectively simulated using Monte Carlo simulations, we use atomistic spin dynamics simulations to model the dynamic properties of Permalloy. The time evolution of atomistic spins is treated by numerical solution of the stochastic Landau-Lifshitz-Gilbert equation with Langevin dynamics applied at the atomistic level \cite{Ellis2015}
\begin{equation}
\frac{\partial \mathbf{S}_i}{\partial t} = -\frac{\gamma }{1 + \lambda^2} \left[\mathbf{S}_i \times \mathbf{B}_{\mathrm{eff}} + \lambda \mathbf{S}_i\times (\mathbf{S}_i \times  \mathbf{B}_{\mathrm{eff}} )\right],
\end{equation}
where $\lambda = 0.0064$ is the microscopic Gilbert damping constant, $\gamma = 1.76 \times 10^{11}$ T$^{-1}$s$^{-1}$ is the absolute value of the gyromagnetic ratio. The effective field $\mathbf{B}_{\mathrm{eff}}$  is calculated from the derivative of the spin Hamiltonian given by Eq.~\ref{eq:ham} with respect to the local spin moment $\mathbf{S}_i$ plus a random thermal Langevin field 
\begin{equation}
\mathbf{B}_{\mathrm{eff}} = -\frac{1}{\mu_i} \frac{\partial \mathscr{H}}{\partial \mathbf{S}_i} + \Gamma(t) \sqrt{\frac{2\lambda k_B \Te}{\gamma \mu_i \Delta t}}
\end{equation}
where $\Gamma$ is a Gaussian distributed 3D random number and $\Te$ is the rescaled electron temperature. As with the equilibrium properties, the classical Heisenberg model fails to reproduce the correct dynamics of real magnetic materials due to the overestimation of the thermal spin fluctuations \cite{EvansPRB2015}. We therefore apply a rescaled temperature to the Langevin thermal field, replicating the quantum statistical effects of the thermal bath on the spin system, where \Te is given by
\begin{equation}
\Te = \Tc \left(\frac{T_{\mathrm{e}}}{\Tc}\right)^{\eta}
\end{equation}
where \Tc is the Curie temperature, $T_{\mathrm{e}}$ is the temperature of the electron bath and $\eta = 1.63$ is the temperature rescaling exponent. Above \Tc no rescaling is applied and the spin system is assumed to follow classical statistics. The temperature rescaling approach corrects for the significant differences in ultrafast dynamics between the classical Heisenberg approach and experimental data for bulk Ni \cite{EvansPRB2015}, and we now utilise the same approach to study ultrafast demagnetization in Permalloy. The time evolution of the electron temperature is simulated using the two-temperature model \cite{Anisimov1974,RaduNature2011,EvansPRB2015} and the dipole fields are computed using a tensor macrocell approximation \cite{Bowden2016} with a 1 nm cell size. The sLLG equation is integrated using a second-order predictor-corrector Heun numerical scheme \cite{EvansVMPR2013} due to its low computational cost and inherent suitability for parallelization using a domain decomposition approach \cite{EvansVMPR2013}.

\begin{figure}[!tb]
\center
\includegraphics[width=8.3cm, trim=0 0 0 0]{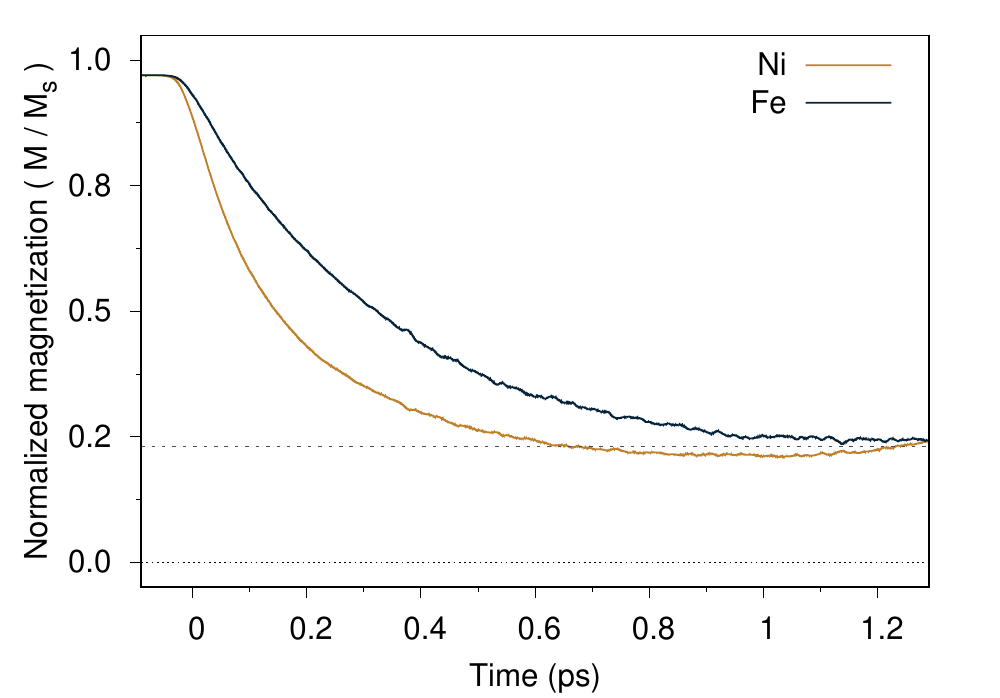}
\caption{(Color online) Simulated ultrafast demagnetization dynamics for Permalloy for a (20 nm)$^3$ representative of a bulk saturated sample. The Fe and Ni sublattices exhibit different dynamics due to different values of magnetic moment, corresponding to approximately the same dissipation rate of angular momentum.}
\label{fig:demag}
\end{figure}

The response of a magnetic material to a laser pulse is often complex with a range of physical effects such as demagnetization\cite{BeaurepairePRL1996,KoopmansNatMat2010,ZhangNANO2018} and switching \cite{StanciuAOS2007, RaduNature2011, OstlerNatCom2012, Mangin2014} but thermal effects have been shown to be the dominant mechanism for demagnetization in ferromagnetic metals  \cite{RothPRX2012,EvansPRB2015,AtxitiaPRB2010} and switching in ferrimagnetic GdFe alloys \cite{OstlerNatCom2012,AtxitiaUFRev2013,YangSCIADV2017}. In two-component ferromagnetic alloys the ultrafast demagnetization dynamics of each magnetic component are different due to different magnetic moments and weakly due to the different microscopic damping  \cite{La-O-VorakiatPRL2009,MathiasPNAS2012,RaduSPIN2015,HinzkePRB2015}. The dependence of the demagnetization time $\tau_{\mathrm{demag}}$ on the magnetic moment $\mu$ and Gilbert damping $\alpha$ reflect the time required to dissipate angular momentum from the spin system to the electrons and lattice, characterized by the approximate proportionality \cite{RaduSPIN2015,EvansAPL2014}
\begin{equation}
    \tau_{\mathrm{demag}} \propto \frac{\mu}{\alpha}.
    \label{eq:tau}
\end{equation}
\begin{figure*}[!tb]
\center
\includegraphics[width=14.3cm, trim=0 0 0 0]{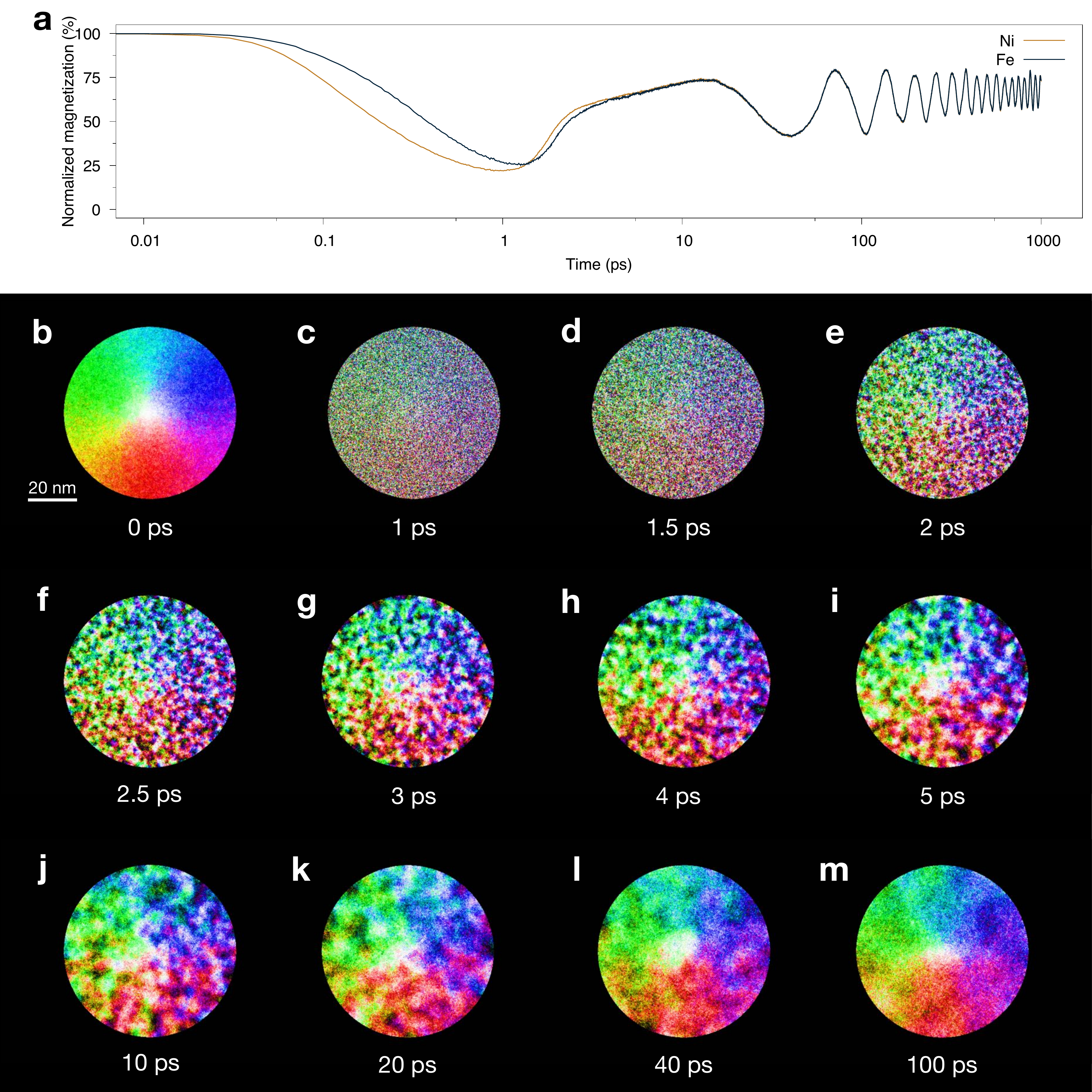}
\caption{(Color online) (a) Simulated ultrafast demagnetization dynamics of the z-component of the magnetization for the Permalloy nanodot normalised to the initial room-temperature value including the vortex configuration. As with the bulk-like sample the demagnetization rate is different for Fe and Ni sublattices, but away from the non-linear dynamics induced by the laser pulse the Ni and Fe sublattices are collinear. After the initial pulse oscillations of the $z$-component of the magnetization are induced and long-lived. (b) Initial thermalised magnetic state of the vortex before the laser excitation is applied. (c-e) Snapshots of the magnetic configuration immediately after the laser excitation showing ultrafast demagnetization and extremely small magnetic textures which induce high frequency spin waves leading to the observed oscillations of the magnetization. (f-l) Time evolution of the magnetic configurations and coalesence of magnetic domains due to magnon transport. (m) Quasi-relaxed magnetic state 100 ps after the laser pulse showing recovery of the vortex state. The non ground-state configuration takes over 1 ns to relax (beyond the timescale of the simulation) to a truly relaxing and thermalise vortex structure.}
\label{fig:dynamics}
\end{figure*}

The question remains whether there is any role of topological magnetic structures on ultrafast demagnetization dynamics, and so for comparison we first model the response of a (10 nm)$^3$ block of Permalloy to a laser pulse with 50 fs width with two-temperature model parameters assumed approximately the same as Nickel as used in \cite{AtxitiaPRB2010}. The system includes periodic boundary conditions and is representative of bulk Ni saturated to a single magnetic domain and equilibrated at $\Te = 300 K$. The short timescale dynamics for the Fe and Ni sublattice are shown in Figure.~\ref{fig:demag}, showing rapid demagnetisation on the sub-picosecond timescale and distinct dynamics for the Fe and Ni sublattices despite their strong exchange coupling. Our results are in close agreement with Radu \textit{et al} \cite{RaduSPIN2015} showing an approximate factor 2 in the characteristic demagnetization time for Fe and Ni in $Ni_{80}Fe_{20}$ Permalloy, although in the case of our simulations the excitation is stronger causing a larger degree of demagnetization.


We now perform the same simulation for a Permalloy nanodot using 960 processors cores with a runtime of approximately 36 hours. The phase diagram for the vortex state depends on the lateral size (radius) and thickness to achieve the appropriate balance of exchange energy and dipole fields \cite{ScholzMAGPAR2003}. Given the computational cost of atomistic calculations, we consider the smallest structure that can support a vortex with a diameter of 70 nm and 20 nm thickness. To generate the vortex magnetic structure we quench the system from a random spin state (infinite temperature) to zero allowing the spins to evolve dynamically to form a ground state configuration with critical damping $(\lambda = 1)$. This naturally forms a vortex state within 100 ps which is fully relaxed for a total of 1 ns of simulation time. The system is then re-thermalized at fixed temperature $\Te = 300 K$. The system is excited by the thermal laser pulse and the dynamic response of the Fe and Ni sublattices is shown in Fig.~\ref{fig:dynamics}(a) with time shown on a logarithmic scale. As for the bulk Permalloy, the Fe and Ni sublattices exhibit different characteristic demagnetization timescales owing to their different magnetic moments. On the 1-10 ps the sublattices partially re-magnetize due to the rapid reduction of the electron temperature due to heat transfer to the lattice. For times greater than 10 ps the Fe and Ni sublattices recover their full collinear character and act as a single ferromagnetic material, while exhibiting oscillations of the perpendicular component of the magnetization for over a nanosecond. 

To gain further insight into the dynamics we plot the magnetic configuration of atoms at the centre plane of the nanodot in Fig.~\ref{fig:dynamics}(b)-(m). The classic vortex pattern is visible in Fig.~\ref{fig:dynamics}(b) before the pulse arrives, superimposed with thermal noise originating from the local random spin fluctuations. We note that a similar effect is seen for Lorentz microscopy images of Permalloy dots and such thermal spin fluctuations could also be considered as a source of noise in such images. One ps after the laser pulse arrives the vortex pattern is disrupted with high frequency noise due to the ultrafast spin wave modes excited by the pulse, shown in Fig.~\ref{fig:dynamics}(c). Over the next few ps the high frequency noise is damped forming a rich domain pattern that evolves quickly in time, consisting of an underlying vortex structure with superimposed noise indicating out-of-plane magnetization components. This evolution is reminiscent of magnon coalescence seen in ferrimagnetic alloys with similarly fast dynamics \cite{IacoccaNatComm2019}. Additionally we see the formation of propagating ultrafast edge spin wave modes (See supplementary movie \S1) that rapidly form and continue for hundreds of picoseconds after the laser pulse. The edge spin waves are the origin of the long-lived oscillations of the perpendicular magnetization component after the laser pulse. The strength of the oscillations decays slowly, likely over several nanoseconds due to the low damping of Permalloy. Nevertheless the underlying vortex state is clearly visible after 100 ps, showing recovery back to the initial vortex magnetization state. Comparison with recent experimental measurements of vortex core demagnetization \cite{RubianodaSilvaPRX2018} show comparable results, with a strong demagnetization of the central vortex region and fast recovery to a new equilibrium value on the sub 10 ps timescale. For larger dots thermal excitation is known to induce multiple vortices \cite{FuSciAdv2018}, but structures of such size are not yet computationally accessible with atomistic simulations.

\begin{figure}[!tb]
\center
\includegraphics[width=8.3cm, trim=0 0 0 0]{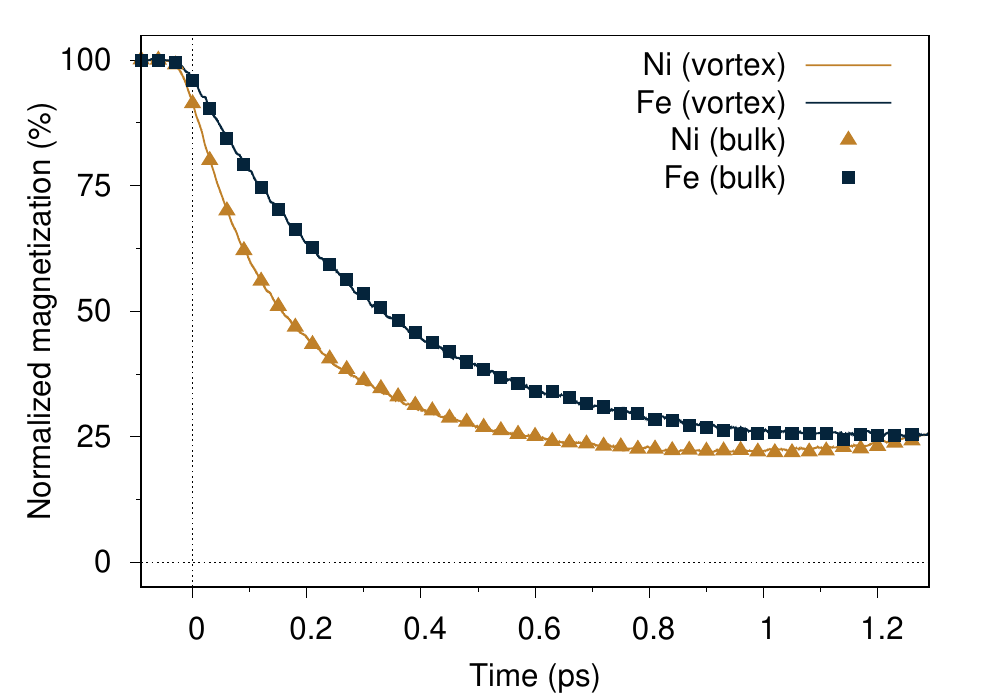}
\caption{(Color online) Comparative simulated demagnetization of Ni and Fe sublattices of Permalloy in bulk and nanodot configurations normalized to the perpendicular magnetization before the pulse. The data show almost exact agreement, demonstrating that topological magnetic structures have no perceptible effect on ultrafast demagnetization, and that the characteristic demagnetisation time is characteristic of atomic scale magnetic properties.}
\label{fig:comp}
\end{figure}

Finally we compare the initial demagnetization for the small bulk-like sample and vortex sample of Py, shown in Fig.~\ref{fig:comp}. The magnetization is normalized to the initial value of the perpendicular magnetization, which is around 95\% for the bulk-like sample and 12.5\% for the nanodot including the vortex. Surprisingly the data is almost exactly the same for both cases, both in terms of the demagnetization time and also the behaviour of the individual magnetic sublattices. This suggests that the relationship between the demagnetisation time and the magnetic moment in Eq.~\ref{eq:tau} is applicable \textit{only} at the atomic scale, and that macroscopic magnetic textures play little role in magnetic processes on the sub-picosecond timescale.

In conclusion, we have applied large-scale massively parallel atomistic spin dynamics simulations to study the ultrafast dynamics of a magnetic vortex subject to an ultrafast thermal laser pulse. We find that the vortex structure is stable for even strong laser  excitation, returning to a clear vortex state after only 5 ps. We also find that the characteristic demagnetization time is unaffected by topological magnetic structures which has important consequences for the manipulation of domain walls and Skyrmions by purely thermal means. More generally, large scale atomistic spin dynamics models provide unprecedented detail for studying the underlying dynamic processes and are approaching lengthscales and timescales accessible with complementary experimental techniques. This presents an exciting opportunity to study the dynamics of complex magnetic structures and devices such as Skyrmions and domain walls including thermal effects beyond the capabilities of the continuum micromagnetic approximation. A particularly interesting topic could be the formation and evolution of Bloch points which are impossible to properly resolve with a continuum approximation but are currently at the limit of computational power available today. The continuing increase in availability of general purpose computing power will likely make the calculations presented here possible on a desktop computer within the next 10-20 years.

\section*{acknowledgments}
This work used the ARCHER UK National Supercomputing Service (http://www.archer.ac.uk) using code enhancements implemented and funded under the ARCHER embedded CSE programme (eCSE0709 and eCSE1307). Data analysis was performed using the \textsc{viking} supercomputing cluster provided by the University of York.

\appendix

\bibliography{library,local}

\clearpage

\end{document}